\documentclass[prl,twocolumn,superscriptaddress,letterpaper,nofootinbib,nopreprintnumbers,noshowpacs]{revtex4-1}

\usepackage{amsmath}
\usepackage{graphicx}% Include figure files
\usepackage{subfigure}
\usepackage{dcolumn}% Align table columns on decimal point
\usepackage{bm}% bold math
\usepackage{hyperref}% add hypertext capabilities
\usepackage{epsfig}
\usepackage{color}
\usepackage[utf8]{inputenc}
\usepackage{comment}
\usepackage[all]{hypcap}
\usepackage{subfiles}
\usepackage{etoolbox}
\usepackage{amsmath}
\usepackage{amssymb}
\newcommand{\ObsTwoStart}{November 30, 2016}
\newcommand{\ObsTwoEnd}{August 25, 2017}
\newcommand{\LiveTime}{117.53 days}

\newcommand{\BankMasses}{$0.19 - 2.0 \thinspace M_\odot$}

\newcommand{\BankMatch}{97\%}
\newcommand{\BankSize}{$992\thinspace461$}

\newcommand{\MaxSpin}{$0.1$}
\newcommand{\FRange}{45--1024 \thinspace Hz}
\newcommand{\SNRLossFromFlow}{8\%}

\newcommand{\LoudestFAR}{3.25 per year}
\newcommand{\NSims}{$913\thinspace931$}
\newcommand{\ZeroPointTwoRate}{\ensuremath {3.7 \times 10^{5} \,\thinspace \mathrm{Gpc}^{-3} \mathrm{yr}^{-1}}}
\newcommand{\OneRate}{\ensuremath {5.2 \times 10^3 \, \thinspace \mathrm{Gpc}^{-3} \mathrm{yr}^{-1}}}
\newcommand{\DMConstraintPointTwo}{16\%}
\newcommand{\DMConstraintOne}{2\%}

\newcommand{\compact}{ultracompact}
\newtoggle{fullauthorlist}
\toggletrue{fullauthorlist}
\newtoggle{endauthorlist}
\toggletrue{endauthorlist}
\date{\today}
\begin{document}
%\linenumbers

\title{Search for sub-solar mass ultracompact binaries in Advanced LIGO’s second observing run}% Force line breaks with \\

\iftoggle{endauthorlist}{
        %
        % Put the author list at the end of the document.
        % Save author, affiliation, and maketitle commands.     
        %
        \let\mymaketitle\maketitle
        \let\myauthor\author
        \let\myaffiliation\affiliation
    %\author{B. P. Abbot \textit{et al.}}
        %\author{The LIGO Scientific Collaboration and The Virgo Collaboration}
        \author{The LIGO Scientific Collaboration}
        %\affiliation{LSC}
        %\email{lsc-spokesperson@ligo.org}
        \author{The Virgo Collaboration}
        %\affiliation{Virgo}
        %\email{virgo-spokesperson@ego-gw.it}
}{
        %
        % Keep the author list on the initial title page.
        %
        \iftoggle{fullauthorlist}{
                %% LSC authorlist in PRD format
%\documentclass[aps,prd,superscriptaddress]{revtex4-1}
%\begin{document}

%\title{LSC February 2019 and Virgo February 2019 author list---LIGO-M1900040\\
%3/26/2019. PRD style}

\author{B.~P.~Abbott}
\affiliation{LIGO, California Institute of Technology, Pasadena, CA 91125, USA}
\author{R.~Abbott}
\affiliation{LIGO, California Institute of Technology, Pasadena, CA 91125, USA}
\author{T.~D.~Abbott}
\affiliation{Louisiana State University, Baton Rouge, LA 70803, USA}
\author{S.~Abraham}
\affiliation{Inter-University Centre for Astronomy and Astrophysics, Pune 411007, India}
\author{F.~Acernese}
\affiliation{Dipartimento di Farmacia, Universit\`a di Salerno, I-84084 Fisciano, Salerno, Italy}
\affiliation{INFN, Sezione di Napoli, Complesso Universitario di Monte S.Angelo, I-80126 Napoli, Italy}
\author{K.~Ackley}
\affiliation{OzGrav, School of Physics \& Astronomy, Monash University, Clayton 3800, Victoria, Australia}
\author{C.~Adams}
\affiliation{LIGO Livingston Observatory, Livingston, LA 70754, USA}
\author{R.~X.~Adhikari}
\affiliation{LIGO, California Institute of Technology, Pasadena, CA 91125, USA}
\author{V.~B.~Adya}
\affiliation{OzGrav, Australian National University, Canberra, Australian Capital Territory 0200, Australia}
\author{C.~Affeldt}
\affiliation{Max Planck Institute for Gravitational Physics (Albert Einstein Institute), D-30167 Hannover, Germany}
\affiliation{Leibniz Universit\"at Hannover, D-30167 Hannover, Germany}
% ...

\collaboration{The LIGO Scientific Collaboration and the Virgo Collaboration}

\author{S.~Shandera}
\affiliation{The Pennsylvania State University, University Park, PA 16802, USA}

%\begin{abstract}
%This is the LSC February 2019 and Virgo February 2019 author list---LIGO-M1900040.
%\end{abstract}

\maketitle

%\end{document}

        }{
                \author{The LIGO Scientific Collaboration}
                \affiliation{LSC}
                \email{lsc-spokesperson@ligo.org}
                \author{The Virgo Collaboration}
                \affiliation{Virgo}
                \email{virgo-spokesperson@ego-gw.it}
        }
}

%\date{\today}

\begin{abstract}
\iftoggle{endauthorlist}{
  % No page break needed
}{
  \newpage
}

%%%%%%%%%%%%%%%%%%%%%%%%%%%%%%%%%%%%%%%%%%%%%%%%%%%%%%%%%%%%%%%
%%%%%%%%%%%%%%%%%%%%%%%%%%%%%%%%%%%%%%%%%%%%%%%%%%%%%%%%%%%%%%%

We present an Advanced LIGO and Advanced Virgo search for sub-solar mass
ultracompact objects in data obtained during Advanced LIGO's second observing
run. In contrast to a previous search of Advanced LIGO data from the first
observing run, this search includes the effects of component spin on the
gravitational waveform. We identify no viable gravitational wave candidates
consistent with sub-solar mass ultracompact binaries with at least one component between
$0.2 - 1.0 \thinspace M_\odot$. We use the null result to constrain the binary merger
rate of ($0.2 \thinspace M_\odot , \, 0.2 \thinspace M_\odot$) binaries to be
less than \ZeroPointTwoRate{} and the binary merger rate of ($1.0 \thinspace
M_\odot , \, 1.0 \thinspace M_\odot$) binaries to be less than \OneRate{}.
Sub-solar mass ultracompact objects are not expected to form via known stellar
evolution channels, though it has been suggested that primordial density
fluctuations or particle dark matter with cooling mechanisms and/or nuclear
interactions could form black holes with sub-solar masses. Assuming a
particular primordial black hole formation model, we constrain a population of
merging $0.2 \thinspace M_\odot$ black holes to account for less than
\DMConstraintPointTwo{} of the dark matter density and a population of merging $1.0
\thinspace M_\odot$ black holes to account for less than \DMConstraintOne{} of
the dark matter density. We discuss how constraints on the merger rate and dark
matter fraction may be extended to arbitrary black hole population models that
predict sub-solar mass binaries.

\end{abstract}

\maketitle

\section{Introduction}

Gravitational wave and multi-messenger astronomy progressed remarkably in
Advanced LIGO~\cite{TheLIGOScientific:2014jea} and Advanced
Virgo's~\cite{TheVirgo:2014hva} second observing run, which included the first
observation of gravitational waves from a binary neutron star
merger~\cite{TheLIGOScientific:2017qsa} and seven of the ten observed
binary black hole mergers~\cite{Abbott:2017vtc,
Abbott:2017gyy,Abbott:2017oio, o1o2catalogpaper}. These
detections, as well as the candidates presented in the gravitational wave
transient catalog (GWTC-1)~\cite{o1o2catalogpaper}, have led to a better
understanding of the populations of compact binaries detectable by ground based
interferometers~\cite{LIGOScientific:2018jsj}. These
observations, however, represent just a portion of the parameter space
that Advanced LIGO and Advanced Virgo currently
search~\cite{DalCanton:2017ala,2018arXiv181205121M} and are sensitive
to~\cite{Magee:2018opb}. We report on an extension of the searched parameter
space in data obtained during O2 to binaries with component masses $< 1
\thinspace M_\odot$. This is the second search for sub-solar mass ultracompact
objects in Advanced LIGO data and the fourth since initial
LIGO~\cite{Abbott:2005pf,Abbott:2007xi,Abbott:2018oah}, as well as the first
search to incorporate spin effects in the modeling of the gravitational wave emission.

There is no known mechanism for the formation of compact objects with masses
well below a solar mass within the standard model of particle physics and the
standard $\Lambda$CDM model of cosmology.  Black holes must be heavier than the
Chandrasekhar limit ($\sim 1.4 \thinspace M_{\odot}$, set by the mass of the
proton \cite{Chandrasekhar:1931ih}) while neutron stars are expected to have
masses greater than about $0.9 \thinspace M_{\odot}$ \cite{kapermeasuringNS,
strobel2001minimum}. A detection of a sub-solar mass object in a merger would
therefore be a clear signal of new physics. Indeed, there are several proposals
that link sub-solar mass compact objects to proposals for the nature of dark
matter, which makes up nearly 85\% of the matter in the Universe. One
possibility is that black holes with masses accessible to ground based interferometers could have
formed deep in the radiation era from the prompt collapse of large primordial
over-densities on the scale of the early time Hubble volume
\cite{Carr:1975qj,nakamura1997gravitational}. The size and abundance of any
such primordial black holes depends on the spectrum of primordial perturbations
and on the equation of state of the early universe \cite{Jedamzik:1996mr,
Widerin:1998my, Georg:2017mqk, Byrnes:2018clq}. An alternative inflationary
mechanism proposes that vacuum bubbles nucleated during inflation may result in
black holes (with masses that can be around a solar mass) after inflation ends
\cite{Deng:2017uwc}.

A different class of possibilities, explored more recently, is motivated by
ideas for the particle nature of dark matter. For example, dark matter may have
a sufficiently complex particle spectrum to support cooling mechanisms that
allow dense regions to collapse into black holes at late times, in processes
analogous to known astrophysical processes \cite{Shandera:2018xkn}.
Alternatively, dark matter may have interactions with nuclear matter that allow
it to collect inside of neutron stars and trigger their collapse to black holes
\cite{Kouvaris:2010jy, deLavallaz:2010wp,PhysRevD.40.3221, Bramante:2015dfa,
Bramante:2014zca,Bramante:2017ulk,Kouvaris:2018wnh}. The details of when dark matter can
collapse a neutron star to form a black hole or other exotic compact object are
still under investigation \cite{Gresham:2018rqo}, but the postulated black
holes will have masses comparable to the progenitor neutron star mass, or
perhaps smaller if some matter can be expelled by rapid rotation of the star
during collapse.

A detection of a sub-solar mass black hole would have far-reaching
implications. In the primordial black hole scenario, the mass and abundance of
the black holes would constrain a combination of the spectrum of initial
density perturbations on very small scales and the equation of state of the
Universe at a time when the typical mass inside a Hubble volume was of the
order of the black hole mass. For particle dark matter scenarios, the abundance
of sub-solar mass black holes would provide a direct estimate of the cooling
rate for dark matter. The black hole mass would constrain the masses of
cosmologically abundant dark matter particles through, for example, the
Chandrasekhar relation for fermions~\cite{Shandera:2018xkn} or analogous
relations for non-interacting
bosons~\cite{1984PhLB..140..329B,UrenaLopez:2002gx}. In the case that all black
holes are observed to be near but not below the mass of neutron stars, the abundance of such
objects would constrain the dark matter-nucleon interaction strength, as well
as the dark matter self-interaction strength and mass(es)
\cite{Kouvaris:2018wnh}.

This letter reports the results of a search for gravitational waves from
sub-solar mass ultracompact binaries using data from Advanced LIGO's second
observing run. No significant candidates consistent with a sub-solar mass
binary were identified. The null result places the tightest constraints to date
on the merger rate and abundance of sub-solar mass ultracompact binaries. We
describe an extension of our merger rate constraints to arbitrary populations
and models under the assumption that the horizon distance controls the
sensitivity of the search. We once more consider the merger rate constraints in
the context of merging primordial black hole populations contributing to the
dark matter~\cite{Abbott:2018oah}. We describe how to extend the dark matter
fraction parameterization to other models by separating LIGO observables from
model dependent quantities. Finally, we conclude with a discussion of the
implications of this search.

\section{Search}
  
We analyze data obtained from \ObsTwoStart{} to \ObsTwoEnd{} during Advanced
LIGO's second observing run (O2).\footnote{Data from Advanced LIGO's second
observing run is available from the Gravitational Wave Open Science Center with
and without noise sources linearly subtracted:
\url{https://www.gw-openscience.org}} Noise artifacts are linearly subtracted
from the data; this includes strong sinusoidal features in both detectors due
to injected calibration frequencies and the AC power grid, as well as laser
beam jitter in the LIGO-Hanford detector data~\cite{DavisMassinger2018}.
\LiveTime{} of coincident data remain after the application of data quality
cuts~\cite{TheLIGOScientific:2017lwt,Abbott:2016trn,abbott2016binary,Nuttall:2018,BBerger:2018}.
The Advanced Virgo
interferometer completed commissioning and joined Advanced LIGO in August 2017
for $15$ days of triple coincident
observation~\cite{o1o2catalogpaper}, however, we only report on the analysis of
data obtained by the LIGO Hanford and LIGO Livingston interferometers.

The search was conducted using publicly available gravitational-wave analysis
software~\cite{Cannon:2011vi,Messick:2016aqy,Sachdev:2019vvd,gstlal,lalsuite,Ajith:2012mn,Capano:2016dsf}.
The initial stage of the search performed a matched-filter analysis using a
discrete bank of template waveforms generated using the TaylorF2
frequency-domain, post-Newtonian inspiral approximant~\cite{Buonanno:2009}.
The template bank used for this search was designed to recover binaries with
component masses of \BankMasses{} and total masses of $0.4 - 4.0 \thinspace
M_\odot$ in the detector frame with \BankMatch{} fidelity, as
in~\cite{Abbott:2018oah}. The search presented here, however, additionally
includes spin effects in the modeling of the gravitational waveform. The bank
is constructed to recover gravitational waves originating from binaries with
component spins purely aligned or anti-aligned with the orbital angular
momentum, and with dimensionless spin magnitudes of \MaxSpin{} or less. The
inclusion of spin effects required denser placement of the waveforms in the
template bank; the resulting bank had \BankSize{} templates, which is nearly
twice as large as the non-spinning bank used in~\cite{Abbott:2018oah}.

In order to reduce the computational burden, matched filtering was only
performed for a subset of Advanced LIGO's full sensitive
band~\cite{Magee:2018opb}. The choice to only analyze the \FRange{} band led to a
detector averaged signal-to-noise ratio (SNR) loss of \SNRLossFromFlow{} when compared to the full
$\sim 10 - 2048 \mathrm{Hz}$ frequency band. This estimated SNR loss is a
property of Advanced LIGO's noise curves and is independent of the templates
used in the search; the discrete nature of the template bank causes an
additional $\lesssim 3\%$ loss in SNR.

Gravitational-wave candidates that were found coincident in both the Hanford
and Livingston detectors were ranked using the logarithm of the
likelihood-ratio ,
$\mathcal{L}$~\cite{Cannon:2011vi,Messick:2016aqy,Sachdev:2019vvd}. For a
candidate with a likelihood-ratio of $\mathcal{L}^*$, we assign a
false-alarm-rate of
\begin{equation} \mathrm{FAR}(\log\mathcal{L}^*)=\frac{N}{T} P\left( \log
\mathcal{L} \ge \log \mathcal{L}^* | \mathrm{noise}\right) , \end{equation}
where $N$ is the number of observed candidates, $T$ is the total live time of
the experiment, and $P\left(\log \mathcal{L} \ge \log \mathcal{L}^* |
\mathrm{noise}\right)$ describes the probability that noise produces a
candidate with a ranking statistic at least as high as the candidate's.

The search recovered the previously detected signal
GW170817~\cite{TheLIGOScientific:2017qsa}, which was observed along with an
electromagnetic counterpart~\cite{GBM:2017lvd}. This signal is consistent with
a binary neutron star. No other viable gravitational wave candidates were
identified. The next loudest candidate was consistent with noise and had a FAR of
\LoudestFAR{}.

\section{Constraint on binary merger rate}
  
\begin{figure} \includegraphics[width=\columnwidth]{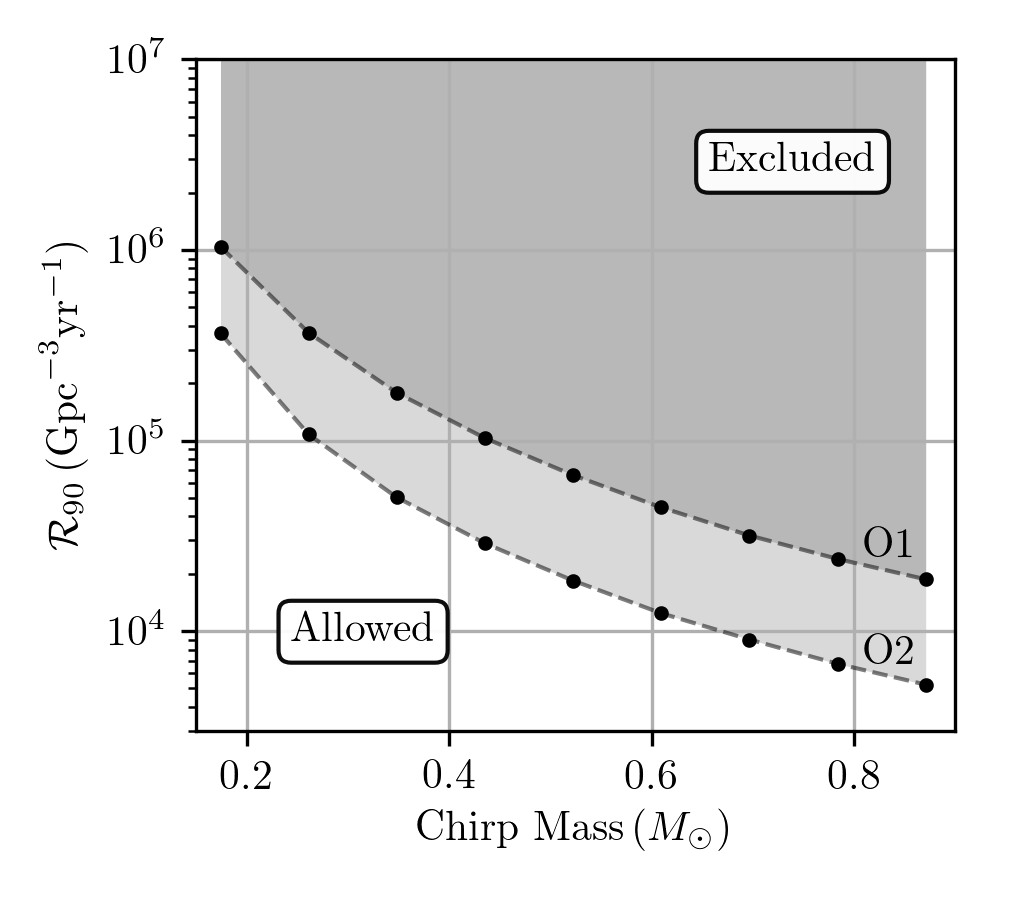}
\caption{\label{fig:rates} The constraint on the merger rate density for equal mass
binaries as a function of chirp mass. The two sets of lines show the
constraints for the O1 search~\cite{Abbott:2018oah} and the O2 search presented
here. The null result from O2 places bounds that are $\sim 3$ times tighter
than the O1 results. The majority of this improvement is due to the increased
coincident observing time in Advanced LIGO's second observing run ($\sim 118$
days vs. $\sim 48$ days), though the improved sensitivity of the detectors led
to an observed physical volume up to $ \sim 50 \%$ larger than in O1 for sub-solar mass ultracompact binaries. } \end{figure}

As in~\cite{Abbott:2018oah}, we consider nine populations of equal mass,
non-spinning binaries that are delta-function distributed in mass, i.e. $m_i
\in \{0.2, 0.3, \dots, 1.0\}$. We injected \NSims{} fake signals into our data;
the injections were randomly oriented and spaced uniformly in distance and
isotropically across the sky. The recovered signals provide an estimate of the
pipeline's detection efficiency as a function of source distance for each equal
mass population. This in turn allows us to estimate the sensitive volume-time
accumulated for each mass bin. We once more use the loudest event statistic
formalism~\cite{Biswas:2007ni} to estimate the upper limit on the binary merger
rate to 90\% confidence,
\begin{equation}\mathcal{R}_i = \frac{2.3}{\langle VT \rangle_i} \end{equation}
These upper limits are shown for equal mass binaries and as a function of chirp
mass in Fig.~\ref{fig:rates}. Although our template bank includes systems with total mass up to $4\thinspace M_\odot$, we only place bounds on the merger rate of systems where both components are $\leq 1 \thinspace M_\odot$. We estimate that detector calibration
uncertainties~\cite{o1o2catalogpaper,2017PhRvD..96j2001C,Viets:2017yvy} and
Monte Carlo errors lead to an uncertainty in our rate constraint of no more
than $20\%$.

Advanced LIGO and Virgo's horizon distance scales as:
\begin{align} D_{\mathrm{horizon}} \propto
\mathcal{M}^{5/6} \sqrt{\int_{f_{\mathrm{min}}}^{f_{\mathrm{max}}}\frac{f^{-7/3}}{S_n(f)} \, df}
\end{align}
where $S_n(f)$ is the noise spectra of the detector and $f_{\mathrm{min}}$ and
$f_{\mathrm{max}}$ are $45$ Hz and $1024$ Hz, respectively.\footnote{The
waveform model used to generate our template bank, TaylorF2, truncates the
waveform at an upper frequency $f_\mathrm{ISCO}$, which corresponds to
radiation from the innermost stable circular orbit of a black hole binary with
mass $M_{\mathrm{total}}$. This frequency is above $f_{\mathrm{max}}$ for all
non-spinning waveforms in our template bank and so does not impact
$D_{\mathrm{horizon}}$.} For a null result, we therefore expect
$\mathcal{R}(\mathcal{M}) \propto \mathcal{M}^{-15/6}$ provided the horizon distance
controls the sensitivity of the search. The observed power law dependence of
the rate constraint on the chirp mass is within $\sim 4\%$ of the expected
$\mathcal{M}^{-15/6}$ dependence; this is well within the error bound on the
rate upper limit and is strong evidence that the chirp mass is the primary
parameter that dictates the sensitivity of the search. Therefore our upper
limits from equal mass systems also apply to unequal mass systems within the
range of mass ratios we have searched over. For verification, we performed a
small injection campaign over five days of coincident data with injected
component masses distributed between $0.19\thinspace M_\odot$ and
$2.0\thinspace M_\odot$ with at least one component $< 1.0 \thinspace M_\odot$. The search sensitivity remained a function of the
chirp mass; this implies that the rate constraints found from the equal mass
injection sets can therefore be applied to systems with arbitrary mass ratios
provided that both component masses lie within $0.20\thinspace M_\odot$ and
$1.0\thinspace M_\odot$ where our injection sets were performed.

The Advanced LIGO and Virgo rate upper limit can be expanded as:
\begin{align} \label{eq:constraint}\mathcal{R}(\mathcal{M}_1, \mathcal{M}_2) =
\int_{\mathcal{M}_1}^{\mathcal{M}_2}\mathcal{R}(\mathcal{M})
\times \psi(\mathcal{M}) \, \mathrm{d}\mathcal{M}\end{align}
where $\mathcal{R}$ is the rate density as a function of chirp mass and
$\psi(\mathcal{M})$ denotes the black hole population distribution in chirp
mass. We ignore the effects of redshift due to the small detector range for sub-solar
mass binaries. Setting $\psi(\mathcal{M}) = \delta(\mathcal{M})$ then reveals
the form of the LIGO constraining rate density, $\mathcal{R}(\mathcal{M})$,
which is shown in Figure~\ref{fig:rates}.
For a given model, $\psi(\mathcal{M})$, $\mathcal{R}(\mathcal{M}_1,
\mathcal{M}_2)$ provides the LIGO rate constraint on that model for chirp
masses between $\mathcal{M}_1$ and $\mathcal{M}_2$. The resulting rate
constraints allow direct comparison of sub-solar mass ultracompact object
models with LIGO observations.

\section{General constraints on sub-solar mass black hole dark matter}
  
\begin{figure} \includegraphics[width=\columnwidth]{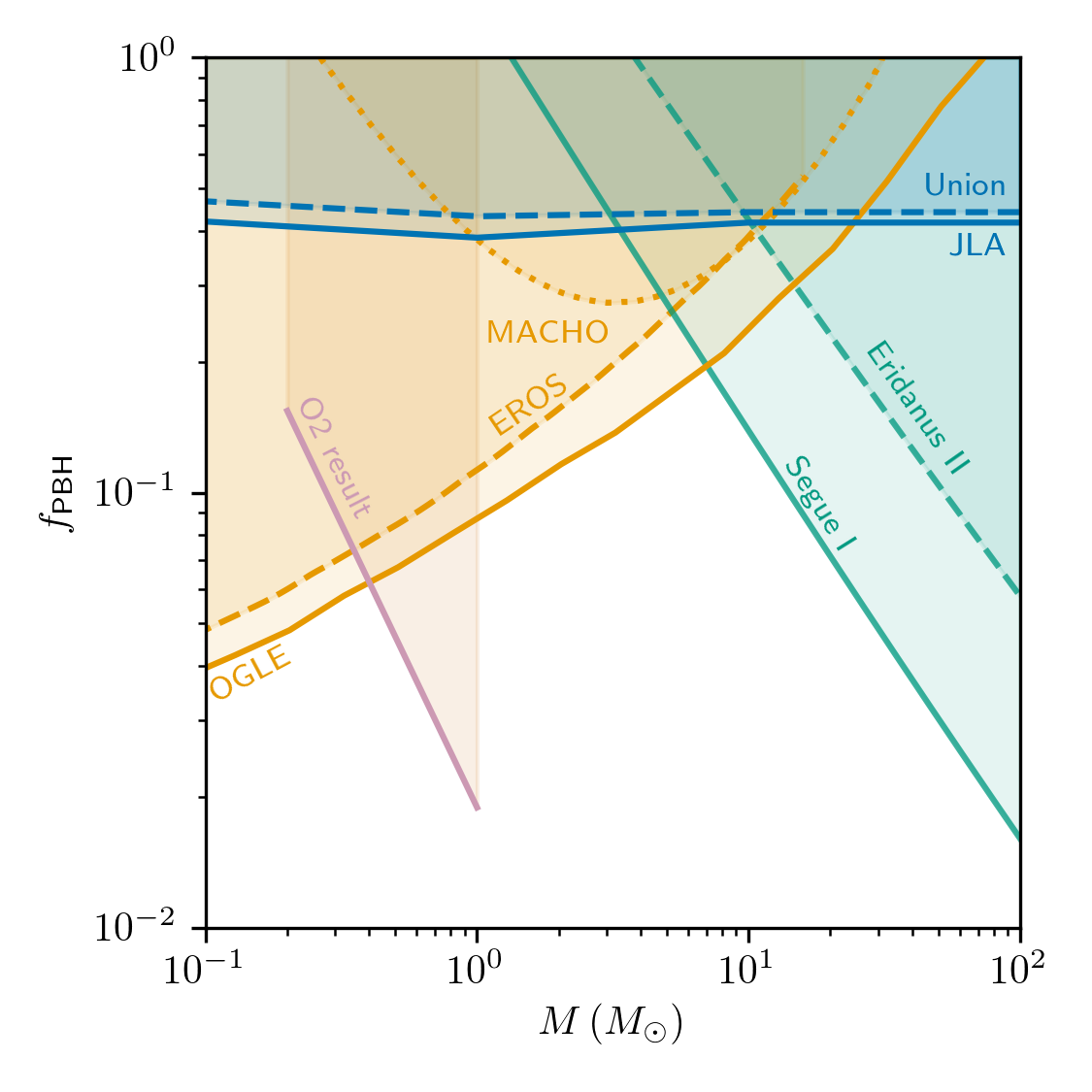}
\caption{\label{fig:fraction} Constraints on the fraction of dark matter
comprised of delta-function distributions of primordial black holes ($ f_{\mathrm{PBH}} =
\rho_{\mathrm{PBH}} / \rho_{\mathrm{DM}} $). Shown here are
(pink) Advanced LIGO constraints from the O2 ultracompact binary search presented here
(solid), (orange) microlensing
constraints provided by the OGLE (solid), EROS
(dashed)~\cite{Tisserand:2006zx}, and MACHO (dotted)
collaborations~\cite{Allsman:2000kg}, (cyan) dynamical constraints from
observations of Segue I (solid)~\cite{Koushiappas:2017chw} and Eridanus II
(dashed)~\cite{Brandt:2016aco} dwarf galaxies, and (blue) supernova lensing
constraints from the Joint Light-curve Analysis (solid) and Union 2.1 (dashed)
datasets~\cite{Zumalacarregui:2017qqd}. There is an inherent population model
dependency in each of these constraints. Advanced LIGO and Advanced Virgo
results carry an additional dependence on the binary fraction of the black hole
population. Advanced LIGO and Advanced Virgo results
use the Planck ``TT,TE,EE+lowP+lensing+ext''
cosmology~\cite{Ade:2015xua}.} \end{figure}

We convert our limits on the merger rate of sub-solar mass
\compact{} objects into a constraint on the abundance of primordial black holes
using our fiducial formation model~\cite{Sasaki:2016jop} first developed in~\cite{nakamura1997gravitational,Ioka1998}
and used previously in LIGO analyses~\cite{Abbott:2005pf,Abbott:2018oah}. We
consider a population of equal mass primordial black holes that is created deep
in the radiation era. We model the binary formation via three-body
interactions, though others have considered the full field of tidal
interactions~\cite{Ali-Haimoud:2017rtz}. By equating the model's predicted
merger rate with the merger rate upper limit provided by Advanced LIGO and
Virgo, we can numerically solve for the upper limit on the PBH abundance. These
constraints are shown in Figure~\ref{fig:fraction}.\footnote{The normalization of the PBH distribution used in our fiducial model~\cite{Sasaki:2016jop} differs by a factor of 2 from the normalization in~\citep{nakamura1997gravitational}. As such, our fiducial model (used here and in~\citep{Abbott:2018oah}) predicts a more conservative PBH merger rate and leads to less constraining limits on $f_{\mathrm{PBH}}$ than would be attained using the model of~\citep{nakamura1997gravitational}.}

This interpretation is highly model dependent; the mass distribution, binary
fraction, and binary formation mechanisms all have a large effect on the
expected present day merger rate and consequently the bounds on the primordial
black hole composition of the dark matter. The Advanced LIGO and Virgo
observables can be separated from the model dependent terms:
\begin{equation}\label{eq:fraction} f_{\mathrm{CO}} = \frac{\rho_{\mathrm{lim}}}{\rho_{\mathrm{CDM}}} \times \frac{1}{f_{\mathrm{obs}}} = \frac{ \mathcal{R}(M) \, T_{\mathrm{obs}} \, M} {\rho_{\mathrm{CDM}} } \times \frac{1}{f_{\mathrm{obs}}} . \end{equation}
Here we use $f_{\mathrm{CO}}$ to refer to the dark matter fraction in
ultracompact objects instead of $f_{\mathrm{PBH}}$ to emphasize that this is
generally applicable to other compact object models that could contribute to
the dark matter~\cite{Shandera:2018xkn}, and not just PBHs. The first term,
$\rho_{\mathrm{lim}} / \rho_{\mathrm{CDM}}$, represents the upper limit on the
fraction of the dark matter contained in presently merging sub-solar mass
ultracompact binaries. In the second term, $f_{\mathrm{obs}}$ describes the
fraction of sub-solar mass ultracompact objects that are observable by Advanced
LIGO and Virgo for a particular model. This is set by the binary fraction and
the probability density of binaries merging at present day. Note that the
merger rate density must be converted from a function of chirp mass to total
mass; this can be done by mapping to total mass for each mass ratio on an equal
chirp mass curve.

Equation~\eqref{eq:fraction} applies to any dark matter model that predicts the
formation of dark compact objects. The abundance of those dark compact objects
can then be expressed as a fraction of the dark matter density.

\section{Conclusion}
  
We have presented the second Advanced LIGO and Advanced Virgo search for
sub-solar mass ultracompact objects. No unambiguous sub-solar mass
gravitational wave candidates were identified. The null result allows us to
place tight constraints on the abundance of sub-solar mass ultracompact
binaries.

This work represents an expansion of previous initial and Advanced LIGO and
Advanced Virgo sub-solar mass searches. First, we have broadened the searched
parameter space to increase sensitivity to systems with non-negligible
component spins. Second, we have presented a method to extend our constraints
on the binary merger rate to arbitrarily distributed populations that contain sub-solar mass ultracompact objects. Combined with
the existing rate limits, this may already be enough to begin constraining
collapsed particulate dark matter models~\cite{Shandera:2018xkn} or the cross
section of nuclear interactions~\cite{Kouvaris:2010jy,
deLavallaz:2010wp,PhysRevD.40.3221, Bramante:2015dfa,
Bramante:2014zca,Kouvaris:2018wnh}. Finally, we have provided a method to
separate Advanced LIGO and Advanced Virgo observables from model dependent terms in our
interpretation of the limits on primordial black hole dark matter.

Ground based interferometer searches for sub-solar mass ultracompact objects
will continue to inform cosmological and particle physics scenarios.  Advanced
LIGO and Advanced Virgo have begun a year long observing run in early
2019, with improved sensitivities~\cite{Aasi:2013wya}. Advanced Virgo will have
more coincident time with the Advanced LIGO detectors over its next observing
run, which will improve network sensitivity and aid in further constraining the
above scenarios.

\section{Acknowledgments}
The authors gratefully acknowledge the support of the United States National
Science Foundation (NSF) for the construction and operation of the LIGO
Laboratory and Advanced LIGO as well as the Science and Technology Facilities
Council (STFC) of the United Kingdom, the Max-Planck-Society (MPS), and the
State of Niedersachsen/Germany for support of the construction of Advanced LIGO
and construction and operation of the GEO600 detector.  Additional support for
Advanced LIGO was provided by the Australian Research Council.  The authors
gratefully acknowledge the Italian Istituto Nazionale di Fisica Nucleare
(INFN), the French Centre National de la Recherche Scientifique (CNRS) and the
Foundation for Fundamental Research on Matter supported by the Netherlands
Organisation for Scientific Research, for the construction and operation of the
Virgo detector and the creation and support  of the EGO consortium.  The
authors also gratefully acknowledge research support from these agencies as
well as by the Council of Scientific and Industrial Research of India, the
Department of Science and Technology, India, the Science \& Engineering
Research Board (SERB), India, the Ministry of Human Resource Development,
India, the Spanish  Agencia Estatal de Investigaci\'on, the Vicepresid\`encia i
Conselleria d'Innovaci\'o, Recerca i Turisme and the Conselleria d'Educaci\'o i
Universitat del Govern de les Illes Balears, the Conselleria d'Educaci\'o,
Investigaci\'o, Cultura i Esport de la Generalitat Valenciana, the National
Science Centre of Poland, the Swiss National Science Foundation (SNSF), the
Russian Foundation for Basic Research, the Russian Science Foundation, the
European Commission, the European Regional Development Funds (ERDF), the Royal
Society, the Scottish Funding Council, the Scottish Universities Physics
Alliance, the Hungarian Scientific Research Fund (OTKA), the Lyon Institute of
Origins (LIO), the Paris \^{I}le-de-France Region, the National Research,
Development and Innovation Office Hungary (NKFI), the National Research
Foundation of Korea, Industry Canada and the Province of Ontario through the
Ministry of Economic Development and Innovation, the Natural Science and
Engineering Research Council Canada, the Canadian Institute for Advanced
Research, the Brazilian Ministry of Science, Technology, Innovations, and
Communications, the International Center for Theoretical Physics South American
Institute for Fundamental Research (ICTP-SAIFR), the Research Grants Council of
Hong Kong, the National Natural Science Foundation of China (NSFC), the
Leverhulme Trust, the Research Corporation, the Ministry of Science and
Technology (MOST), Taiwan and the Kavli Foundation.  The authors gratefully
acknowledge the support of the NSF, STFC, MPS, INFN, CNRS and the State of
Niedersachsen/Germany for provision of computational resources. Computing
resources and personnel for this project were provided by the Pennsylvania
State University. This article has been assigned the document number
LIGO-P1900037.

\newpage

\iftoggle{endauthorlist}{
  %
  % Restore the author, affiliation and maketitle commands.
  %
  \let\author\myauthor
  \date[]{}
  \let\affiliation\myaffiliation
  \let\maketitle\mymaketitle
  \title{Authors}
  \pacs{}
  
  \newpage
  \maketitle
}


\begin{thebibliography}{0}%
\makeatletter
\providecommand \@ifxundefined [1]{%
 \@ifx{#1\undefined}
}%
\providecommand \@ifnum [1]{%
 \ifnum #1\expandafter \@firstoftwo
 \else \expandafter \@secondoftwo
 \fi
}%
\providecommand \@ifx [1]{%
 \ifx #1\expandafter \@firstoftwo
 \else \expandafter \@secondoftwo
 \fi
}%
\providecommand \natexlab [1]{#1}%
\providecommand \enquote  [1]{``#1''}%
\providecommand \bibnamefont  [1]{#1}%
\providecommand \bibfnamefont [1]{#1}%
\providecommand \citenamefont [1]{#1}%
\providecommand \href@noop [0]{\@secondoftwo}%
\providecommand \href [0]{\begingroup \@sanitize@url \@href}%
\providecommand \@href[1]{\@@startlink{#1}\@@href}%
\providecommand \@@href[1]{\endgroup#1\@@endlink}%
\providecommand \@sanitize@url [0]{\catcode `\\12\catcode `\$12\catcode
  `\&12\catcode `\#12\catcode `\^12\catcode `\_12\catcode `\%12\relax}%
\providecommand \@@startlink[1]{}%
\providecommand \@@endlink[0]{}%
\providecommand \url  [0]{\begingroup\@sanitize@url \@url }%
\providecommand \@url [1]{\endgroup\@href {#1}{\urlprefix }}%
\providecommand \urlprefix  [0]{URL }%
\providecommand \Eprint [0]{\href }%
\providecommand \doibase [0]{http://dx.doi.org/}%
\providecommand \selectlanguage [0]{\@gobble}%
\providecommand \bibinfo  [0]{\@secondoftwo}%
\providecommand \bibfield  [0]{\@secondoftwo}%
\providecommand \translation [1]{[#1]}%
\providecommand \BibitemOpen [0]{}%
\providecommand \bibitemStop [0]{}%
\providecommand \bibitemNoStop [0]{.\EOS\space}%
\providecommand \EOS [0]{\spacefactor3000\relax}%
\providecommand \BibitemShut  [1]{\csname bibitem#1\endcsname}%
\let\auto@bib@innerbib\@empty
%</preamble>
\end{thebibliography}%


\begin{thebibliography}{10}

\bibitem{TheLIGOScientific:2014jea}
J.~Aasi et~al.
\newblock {Advanced LIGO}.
\newblock {\em Class. Quant. Grav.}, 32:074001, 2015.

\bibitem{TheVirgo:2014hva}
F.~Acernese et~al.
\newblock {Advanced Virgo: a second-generation interferometric gravitational
  wave detector}.
\newblock {\em Class. Quant. Grav.}, 32(2):024001, 2015.

\bibitem{TheLIGOScientific:2017qsa}
B.~P. Abbott et~al.
\newblock {GW170817: Observation of Gravitational Waves from a Binary Neutron
  Star Inspiral}.
\newblock {\em Phys. Rev. Lett.}, 119(16):161101, 2017.

\bibitem{Abbott:2017vtc}
Benjamin~P. Abbott et~al.
\newblock {GW170104: Observation of a 50-Solar-Mass Binary Black Hole
  Coalescence at Redshift 0.2}.
\newblock {\em Phys. Rev. Lett.}, 118(22):221101, 2017.

\bibitem{Abbott:2017gyy}
B..~P.. Abbott et~al.
\newblock {GW170608: Observation of a 19-solar-mass Binary Black Hole
  Coalescence}.
\newblock {\em Astrophys. J.}, 851(2):L35, 2017.

\bibitem{Abbott:2017oio}
B.~P. Abbott et~al.
\newblock {GW170814: A Three-Detector Observation of Gravitational Waves from a
  Binary Black Hole Coalescence}.
\newblock {\em Phys. Rev. Lett.}, 119(14):141101, 2017.

\bibitem{o1o2catalogpaper}
B.~P. Abbott et~al.
\newblock {GWTC-1: A Gravitational-Wave Transient Catalog of Compact Binary
  Mergers Observed by LIGO and Virgo during the First and Second Observing
  Runs}.
\newblock {\em Preprint: arXiv 1811.12907}, 2018.

\bibitem{LIGOScientific:2018jsj}
B.~P. Abbott et~al.
\newblock {Binary Black Hole Population Properties Inferred from the First and
  Second Observing Runs of Advanced LIGO and Advanced Virgo}.
\newblock 2018.

\bibitem{DalCanton:2017ala}
Tito Dal~Canton and Ian~W. Harry.
\newblock {Designing a template bank to observe compact binary coalescences in
  Advanced LIGO's second observing run}.
\newblock {\em Preprint: arXiv 1705.01845}, 2017.

\bibitem{2018arXiv181205121M}
Debnandini {Mukherjee}, Sarah {Caudill}, Ryan {Magee}, Cody {Messick}, Stephen
  {Privitera}, Surabhi {Sachdev}, Kent {Blackburn}, Patrick {Brady}, Patrick
  {Brockill}, Kipp {Cannon}, Sydney~J. {Chamberlin}, Deep {Chatterjee}, Jolien
  D.~E. {Creighton}, Heather {Fong}, Patrick {Godwin}, Chad {Hanna}, Shasvath
  {Kapadia}, Ryan~N. {Lang}, Tjonnie G.~F. {Li}, Rico K.~L. {Lo}, Duncan
  {Meacher}, Alex {Pace}, Laleh {Sadeghian}, Leo {Tsukada}, Leslie {Wade},
  Madeline {Wade}, Alan {Weinstein}, and Liting {Xiao}.
\newblock {The GstLAL template bank for spinning compact binary mergers in the
  second observation run of Advanced LIGO and Virgo}.
\newblock {\em Preprint: arXiv 1812.05121}, Dec 2018.

\bibitem{Magee:2018opb}
Ryan Magee, Anne-Sylvie Deutsch, Phoebe McClincy, Chad Hanna, Christian Horst,
  Duncan Meacher, Cody Messick, Sarah Shandera, and Madeline Wade.
\newblock {Methods for the detection of gravitational waves from subsolar mass
  ultracompact binaries}.
\newblock {\em Phys. Rev.}, D98(10):103024, 2018.

\bibitem{Abbott:2005pf}
B.~Abbott et~al.
\newblock {Search for gravitational waves from primordial black hole binary
  coalescences in the galactic halo}.
\newblock {\em Phys. Rev.}, D72:082002, 2005.

\bibitem{Abbott:2007xi}
B.~Abbott et~al.
\newblock {Search for gravitational waves from binary inspirals in S3 and S4
  LIGO data}.
\newblock {\em Phys. Rev.}, D77:062002, 2008.

\bibitem{Abbott:2018oah}
B.~P. Abbott et~al.
\newblock {Search for Subsolar-Mass Ultracompact Binaries in Advanced LIGO’s
  First Observing Run}.
\newblock {\em Phys. Rev. Lett.}, 121(23):231103, 2018.

\bibitem{Chandrasekhar:1931ih}
Subrahmanyan Chandrasekhar.
\newblock {The maximum mass of ideal white dwarfs}.
\newblock {\em Astrophys. J.}, 74:81--82, 1931.

\bibitem{kapermeasuringNS}
L.~{Kaper}, A.~{van der Meer}, M.~{van Kerkwijk}, and E.~{van den Heuvel}.
\newblock {Measuring the Masses of Neutron Stars}.
\newblock {\em The Messenger}, 126:27--31, December 2006.

\bibitem{strobel2001minimum}
Klaus Strobel and Manfred~K Weigel.
\newblock On the minimum and maximum mass of neutron stars and the delayed
  collapse.
\newblock {\em Astronomy \& Astrophysics}, 367(2):582--587, 2001.

\bibitem{Carr:1975qj}
Bernard~J. Carr.
\newblock {The Primordial black hole mass spectrum}.
\newblock {\em Astrophys. J.}, 201:1--19, 1975.

\bibitem{nakamura1997gravitational}
Takashi Nakamura, Misao Sasaki, Takahiro Tanaka, and Kip~S Thorne.
\newblock Gravitational waves from coalescing black hole macho binaries.
\newblock {\em The Astrophysical Journal Letters}, 487(2):L139, 1997.

\bibitem{Jedamzik:1996mr}
Karsten Jedamzik.
\newblock {Primordial black hole formation during the QCD epoch}.
\newblock {\em Phys. Rev.}, D55:5871--5875, 1997.

\bibitem{Widerin:1998my}
Peter Widerin and Christoph Schmid.
\newblock {Primordial black holes from the QCD transition?}
\newblock {\em Preprint: arXiv astro-ph/9808142}, 1998.

\bibitem{Georg:2017mqk}
Julian Georg and Scott Watson.
\newblock {A Preferred Mass Range for Primordial Black Hole Formation and Black
  Holes as Dark Matter Revisited}.
\newblock {\em JHEP}, 09:138, 2017.

\bibitem{Byrnes:2018clq}
Christian~T. Byrnes, Mark Hindmarsh, Sam Young, and Michael R.~S. Hawkins.
\newblock {Primordial black holes with an accurate QCD equation of state}.
\newblock {\em JCAP}, 1808(08):041, 2018.

\bibitem{Deng:2017uwc}
Heling Deng and Alexander Vilenkin.
\newblock {Primordial black hole formation by vacuum bubbles}.
\newblock {\em JCAP}, 1712(12):044, 2017.

\bibitem{Shandera:2018xkn}
Sarah Shandera, Donghui Jeong, and Henry S.~Grasshorn Gebhardt.
\newblock {Gravitational Waves from Binary Mergers of Subsolar Mass Dark Black
  Holes}.
\newblock {\em Phys. Rev. Lett.}, 120(24):241102, 2018.

\bibitem{Kouvaris:2010jy}
Chris Kouvaris and Peter Tinyakov.
\newblock {Constraining Asymmetric Dark Matter through observations of compact
  stars}.
\newblock {\em Phys. Rev.}, D83:083512, 2011.

\bibitem{deLavallaz:2010wp}
Arnaud de~Lavallaz and Malcolm Fairbairn.
\newblock {Neutron Stars as Dark Matter Probes}.
\newblock {\em Phys. Rev.}, D81:123521, 2010.

\bibitem{PhysRevD.40.3221}
Itzhak Goldman and Shmuel Nussinov.
\newblock Weakly interacting massive particles and neutron stars.
\newblock {\em Phys. Rev. D}, 40:3221--3230, Nov 1989.

\bibitem{Bramante:2015dfa}
Joseph Bramante and Fatemeh Elahi.
\newblock {Higgs portals to pulsar collapse}.
\newblock {\em Phys. Rev.}, D91(11):115001, 2015.

\bibitem{Bramante:2014zca}
Joseph Bramante and Tim Linden.
\newblock {Detecting Dark Matter with Imploding Pulsars in the Galactic
  Center}.
\newblock {\em Phys. Rev. Lett.}, 113(19):191301, 2014.

\bibitem{Bramante:2017ulk}
Joseph Bramante, Tim Linden, and Yu-Dai Tsai.
\newblock {Searching for dark matter with neutron star mergers and quiet
  kilonovae}.
\newblock {\em Phys. Rev.}, D97(5):055016, 2018.

\bibitem{Kouvaris:2018wnh}
Chris Kouvaris, Peter Tinyakov, and Michel H.~G. Tytgat.
\newblock {NonPrimordial Solar Mass Black Holes}.
\newblock {\em Phys. Rev. Lett.}, 121(22):221102, 2018.

\bibitem{Gresham:2018rqo}
Moira~I. Gresham and Kathryn~M. Zurek.
\newblock {Asymmetric Dark Stars and Neutron Star Stability}.
\newblock {\em Phys. Rev.}, D99(8):083008, 2019.

\bibitem{1984PhLB..140..329B}
J.~D. {Breit}, S.~{Gupta}, and A.~{Zaks}.
\newblock {Cold Bose stars}.
\newblock {\em Physics Letters B}, 140:329--332, June 1984.

\bibitem{UrenaLopez:2002gx}
L.~A. Urena-Lopez, T.~Matos, and R.~Becerril.
\newblock {Inside oscillatons}.
\newblock {\em Class. Quant. Grav.}, 19:6259--6277, 2002.

\bibitem{DavisMassinger2018}
D.~Davis, T.J. Massinger, et~al.
\newblock {Improving the Sensitivity of Advanced LIGO Using Noise Subtraction}.
\newblock {\em Preprint: arXiv 1809.05348}, 2018.

\bibitem{TheLIGOScientific:2017lwt}
B~P Abbott et~al.
\newblock {Effects of data quality vetoes on a search for compact binary
  coalescences in Advanced LIGO's first observing run}.
\newblock {\em Class. Quant. Grav.}, 35(6):065010, 2018.

\bibitem{Abbott:2016trn}
B.~P. Abbott et~al.
\newblock {Characterization of transient noise in Advanced LIGO relevant to
  gravitational wave signal GW150914}.
\newblock {\em Class. Quant. Grav.}, 33(13), 2016.

\bibitem{abbott2016binary}
BP~Abbott, R~Abbott, TD~Abbott, MR~Abernathy, F~Acernese, K~Ackley, C~Adams,
  T~Adams, P~Addesso, RX~Adhikari, et~al.
\newblock Binary black hole mergers in the first advanced ligo observing run.
\newblock {\em Physical Review X}, 6(4):041015, 2016.

\bibitem{Nuttall:2018}
L.K. Nuttall.
\newblock {Characterizing transient noise in the LIGO detectors}.
\newblock {\em Royal Society Proceedings A}, 376(2120), 2018.

\bibitem{BBerger:2018}
B.K. Berger.
\newblock {Identification and mitigation of Advanced LIGO noise sources}.
\newblock {\em J. Phys.: Conf. Ser.}, 957(012004), 2018.

\bibitem{Cannon:2011vi}
Kipp Cannon et~al.
\newblock {Toward Early-Warning Detection of Gravitational Waves from Compact
  Binary Coalescence}.
\newblock {\em Astrophys. J.}, 748:136, 2012.

\bibitem{Messick:2016aqy}
Cody Messick et~al.
\newblock {Analysis Framework for the Prompt Discovery of Compact Binary
  Mergers in Gravitational-wave Data}.
\newblock {\em Phys. Rev.}, D95(4):042001, 2017.

\bibitem{Sachdev:2019vvd}
Surabhi Sachdev et~al.
\newblock {The GstLAL Search Analysis Methods for Compact Binary Mergers in
  Advanced LIGO's Second and Advanced Virgo's First Observing Runs}.
\newblock {\em Preprint: arXiv 1901.08580}, 2019.

\bibitem{gstlal}
Gst{LAL} software: git.ligo.org/lscsoft/gstlal.

\bibitem{lalsuite}
{LAL} software: git.ligo.org/lscsoft/lalsuite.

\bibitem{Ajith:2012mn}
P.~Ajith, N.~Fotopoulos, S.~Privitera, A.~Neunzert, and A.~J. Weinstein.
\newblock {Effectual template bank for the detection of gravitational waves
  from inspiralling compact binaries with generic spins}.
\newblock {\em Phys. Rev.}, D89(8):084041, 2014.

\bibitem{Capano:2016dsf}
Collin Capano, Ian Harry, Stephen Privitera, and Alessandra Buonanno.
\newblock {Implementing a search for gravitational waves from binary black
  holes with nonprecessing spin}.
\newblock {\em Phys. Rev.}, D93(12):124007, 2016.

\bibitem{Buonanno:2009}
Alessandra {Buonanno}, Bala~R. {Iyer}, Evan {Ochsner}, Yi~{Pan}, and B.~S.
  {Sathyaprakash}.
\newblock {Comparison of post-Newtonian templates for compact binary inspiral
  signals in gravitational-wave detectors}.
\newblock {\em \prd}, 80:084043, October 2009.

\bibitem{GBM:2017lvd}
B.~P. Abbott et~al.
\newblock {Multi-messenger Observations of a Binary Neutron Star Merger}.
\newblock {\em Astrophys. J.}, 848(2):L12, 2017.

\bibitem{Biswas:2007ni}
Rahul Biswas, Patrick~R. Brady, Jolien D.~E. Creighton, and Stephen Fairhurst.
\newblock {The Loudest event statistic: General formulation, properties and
  applications}.
\newblock {\em Class. Quant. Grav.}, 26:175009, 2009.
\newblock [Erratum: Class. Quant. Grav.30,079502(2013)].

\bibitem{2017PhRvD..96j2001C}
Craig {Cahillane}, Joe {Betzwieser}, Duncan~A. {Brown}, Evan {Goetz}, Evan~D.
  {Hall}, Kiwamu {Izumi}, Shivaraj {Kand hasamy}, Sudarshan {Karki}, Jeff~S.
  {Kissel}, Greg {Mendell}, Richard~L. {Savage}, Darkhan {Tuyenbayev}, Alex
  {Urban}, Aaron {Viets}, Madeline {Wade}, and Alan~J. {Weinstein}.
\newblock {Calibration uncertainty for Advanced LIGO's first and second
  observing runs}.
\newblock {\em \prd}, 96:102001, Nov 2017.

\bibitem{Viets:2017yvy}
Aaron Viets et~al.
\newblock {Reconstructing the calibrated strain signal in the Advanced LIGO
  detectors}.
\newblock {\em Class. Quant. Grav.}, 35(9):095015, 2018.

\bibitem{Sasaki:2016jop}
Misao Sasaki, Teruaki Suyama, Takahiro Tanaka, and Shuichiro Yokoyama.
\newblock {Primordial Black Hole Scenario for the Gravitational-Wave Event
  GW150914}.
\newblock {\em Phys. Rev. Lett.}, 117(6):061101, 2016.

\bibitem{Ioka1998}
Kunihito Ioka, Takeshi Chiba, Takahiro Tanaka, and Takashi Nakamura.
\newblock {Black Hole Binary Formation in the Expanding Universe : Three Body
  Problem Approximation}.
\newblock {\em Physical Review D}, 58(6):063003, 1998.

\bibitem{Ali-Haimoud:2017rtz}
Yacine Ali-Ha{\"\i}moud, Ely~D. Kovetz, and Marc Kamionkowski.
\newblock {Merger rate of primordial black-hole binaries}.
\newblock {\em Phys. Rev.}, D96(12):123523, 2017.

\bibitem{Tisserand:2006zx}
P.~Tisserand et~al.
\newblock {Limits on the Macho Content of the Galactic Halo from the EROS-2
  Survey of the Magellanic Clouds}.
\newblock {\em Astron. Astrophys.}, 469:387--404, 2007.

\bibitem{Allsman:2000kg}
R.~A. Allsman et~al.
\newblock {MACHO project limits on black hole dark matter in the 1-30 solar
  mass range}.
\newblock {\em Astrophys. J.}, 550:L169, 2001.

\bibitem{Koushiappas:2017chw}
Savvas~M. Koushiappas and Abraham Loeb.
\newblock {Dynamics of Dwarf Galaxies Disfavor Stellar-Mass Black Holes as Dark
  Matter}.
\newblock {\em Phys. Rev. Lett.}, 119(4):041102, 2017.

\bibitem{Brandt:2016aco}
Timothy~D. Brandt.
\newblock {Constraints on MACHO Dark Matter from Compact Stellar Systems in
  Ultra-Faint Dwarf Galaxies}.
\newblock {\em Astrophys. J.}, 824(2):L31, 2016.

\bibitem{Zumalacarregui:2017qqd}
Miguel Zumalacarregui and Uros Seljak.
\newblock {Limits on stellar-mass compact objects as dark matter from
  gravitational lensing of type Ia supernovae}.
\newblock {\em Phys. Rev. Lett.}, 121(14):141101, 2018.

\bibitem{Ade:2015xua}
P.~A.~R. Ade et~al.
\newblock {Planck 2015 results. XIII. Cosmological parameters}.
\newblock {\em Astron. Astrophys.}, 594:A13, 2016.

\bibitem{Aasi:2013wya}
Benjamin~P. Abbott et~al.
\newblock {Prospects for Observing and Localizing Gravitational-Wave Transients
  with Advanced LIGO, Advanced Virgo and KAGRA}.
\newblock {\em Living Rev. Rel.}, 21:3, 2018.
\newblock [Living Rev. Rel.19,1(2016)].

\end{thebibliography}
\end{document}